\documentstyle[11pt,epsfig]{article}
\date{}
\begin{document}
\title{{\bf Classical and quantum massive cosmology for the open FRW universe }}
\author{Babak Vakili$^{1}$\thanks{b-vakili@iauc.ac.ir}\,\, and\,\,\
Nima Khosravi$^{2}$\thanks{nima@aims.ac.za}\,\,\,\\\\$^1${\small
{\it Department of Physics, Chalous Branch, IAU, P.O. Box 46615-397,
Chalous, Iran}}\\$^2${\small {\it Cosmology Group, African Institute
for Mathematical Sciences, Muizenberg 7945, Cape Town, South
Africa}} $^{}${}} \maketitle

\begin{abstract}
In an open Friedmann-Robertson-Walker (FRW) space background, we
study the classical and quantum cosmological models in the
framework of the recently proposed nonlinear massive gravity
theory. Although the constraints which are present in this theory
prevent it from admitting the flat and closed FRW models as its
cosmological solutions, for the open FRW universe, it is not the
case. We have shown that, either in the absence of matter or in
the presence of a perfect fluid, the classical field equations of
such a theory adopt physical solutions for the open FRW model, in
which the mass term shows itself as a cosmological constant. These
classical solutions consist of two distinguishable branches: One
is a contacting universe which tends to a future singularity with
zero size, while another is an expanding universe having a past
singularity from which it begins its evolution. A classically
forbidden region separates these two branches from each other. We
then employ the familiar canonical quantization procedure in the
given cosmological setting to find the cosmological wave
functions. We use the resulting wave function to investigate the
possibility of the avoidance of classical singularities due to
quantum effects. It is shown that the quantum expectation values
of the scale factor, although they have either contracting or
expanding phases like their classical counterparts, are not
disconnected from each other. Indeed, the classically forbidden
region may be replaced by a bouncing period in which the scale
factor bounces from the contraction to its expansion eras. Using
the Bohmian approach of quantum mechanics, we also compute the
Bohmian trajectory and the quantum potential related to the
system, which their analysis shows are the direct effects of the
mass term on the dynamics of the universe.
\vspace{5mm}\noindent\vspace{0.8mm}\newline
\end{abstract}
\section{Introduction}
General Relativity (GR) introduced by Einstein began a renaissance
in scientific thought which changed our viewpoint on the concept
of space-time geometry and gravity. The interpretation of
gravitational force as a modification of geometrical structure of
space-time made and makes this force distinguishable from other
fundamental interactions, although there are arguments which
support the idea that the other interactions may also have
geometrical origin. Because of the unknown behavior of
gravitational interaction at short distances, this distinction may
have some roots in the heart of problems with quantum gravity.
Therefore, any hope of dealing with such concepts would be in vain
unless a reliable quantum theory of gravity can be constructed. In
the absence of a full theory of quantum gravity, it would be then
useful to describe its quantum aspects within the context of
modified theories of gravity. From a field theory point of view,
the gravitational force in GR can be represented as a field theory
in which the space-time metric plays the role of the fields and
the particle that is responsible to propagate gravity is named
graviton. Then, naturally in comparison to the other field
theories, one may ask about the different properties of such a
particle. The answer to this question is deduced by the linearized
form of GR and expansion of the space-time metric $g_{\mu\nu}$,
around a fixed background geometry $\eta_{\mu\nu}$, as
$g_{\mu\nu}=\eta_{\mu\nu}+h_{\mu\nu}$, where $h_{\mu\nu}$ is the
field representation of the graviton. Eventually, it is possible
to show that the graviton is a \textit{massless} spin-$2$
particle.

Then, since our knowledge about the behavior of gravity at very
long distances is also incomplete, a question arises: Is it
possible to consider a small nonvanishing mass for the graviton,
i.e., a \textit{massive} spin-$2$ particle? In the first attempts
to deal with this question, it seemed that adding a mass term to
the action may be sufficient. This was done by Fierz and Pauli
\cite{pauli}. However, it was shown that by considering the number
of degrees of freedom, this model suffers from the existence of a
ghost field, the so-called Boulware-Deser ghost \cite{deser},
after studying the non-linear terms. This fact made massive
gravity an abandoned theory for a while. Recently, de Rahm and
Gabadadze proposed a new scenario in which they have shown that it
is possible to have a ghost-free massive gravity even at the
non-linear level \cite{derahm}. That was a positive signal in this
area, and the early results in this subject have been followed by
a number of works that address different aspects of massive
gravity \cite{works}. As in the case of the other modified
theories of gravity, it is important to seek cosmological
solutions in the newly proposed massive theory of gravity. This is
done by the authors of Ref. \cite{Amico}, who show that the
existence of some constraints prevent the theory from having the
nontrivial homogeneous and isotropic cosmological solutions.
Indeed, what is shown in Ref. \cite{Amico} is that, beginning with
the flat FRW ansatz in the context of massive gravity, the
corresponding field equations result in nothing but the Minkowski
metric. However, by re-examination of the conditions, the authors
of Refs. \cite{Emir,Emir1} have shown that, for the open FRW
model, this is not the case, and the nonlinear massive gravity
admits the open FRW as a compatible solution for its field
equations. Another progresses to find the massive cosmologies lie
in the field of the bi-metric theories of gravity; see, for
instance, Ref. \cite{nima} based on the works of Hassan and Rosen
\cite{hassan}, in which they show that a bi-metric representation
for massive gravity exists.

Our purpose in the present paper is to continue the works of the
authors of Ref. \cite{Emir,Emir1} in greater detail, based on the
Hamiltonian formalism of the open FRW cosmology in the framework
of massive gravity. We obtain the solutions to the vacuum and
perfect fluid classical field equations and investigate their
different aspects, such as the roll of the graviton's mass as a
cosmological constant, the appearance of singularities, and the
late time expansion. We then consider the problem at hand in the
context of canonical quantum cosmology to see how the classical
picture will be modified. Our final results show that the singular
behavior of the classical cosmology will be replaced by a bouncing
one when quantum mechanical considerations are taken into account.
This means that the quantization of the model suggests the
existence of a minimal size for the corresponding universe. We
shall also study the quantum model by the Bohmian approach of
quantum mechanics to show how the mass term exhibits its direct
effects on the evolution of the system.

The structure of the paper is as follows. In section 2, we briefly
present the basic elements of the issue of massive gravity and its
canonical Hamiltonian for a given open FRW universe. In section 3,
classical cosmological dynamics is introduced for the vacuum and
perfect fluid. Quantization of the model is the subject of section
4, and in section 5, the Bohmian approach of quantum mechanics is
applied to the model. Finally, the conclusions are summarized in
section 6.

\section{Preliminary set-up}
In this section, we start by briefly studying the nonlinear
massive gravity action presented in Refs. \cite{Emir,Emir1} for
the open FRW model, where the metric is given by
\begin{equation}\label{A}
ds^2=g_{\mu\nu}dx^{\mu}dx^{\nu}=-N^2(t)dt^2+a^2(t)\left[dx^2+dy^2+dz^2-\frac{|K|(xdx+ydy+zdz)^2}{1+|K|(x^2+y^2+z^2)}\right],\end{equation}
with $N(t)$ and $a(t)$ being the lapse function and the scale
factor, respectively, and $K=-1$ denoting the curvature index.
Here we work in units where $c=\hbar=16\pi G=1$. In the massive
gravity scenario one considers a metric perturbation as
\cite{Amico,Clau}
\begin{equation}\label{A1}
g_{\mu\nu}=\eta_{\mu\nu}+h_{\mu\nu}=\eta_{ab}\partial_{\mu}\phi^a(x)\partial_{\nu}\phi^b(x)+H_{\mu\nu},\end{equation}where
$\eta_{ab}=\mbox{diag}(-1,1,1,1)$ and $\phi^a(x)$ are four scalar
fields known as St\"{u}ckelberg scalars and are introduced to keep
the principle of general covariance also in massive general
relativity \cite{Arkani}. It is clear that the first term in
(\ref{A1}) is a representation of the Minkowski space-time in
terms of the coordinate system $(\phi^0,\phi^i)$ and thus the
tensor $H_{\mu\nu}$ is responsible for describing the propagation
of gravity in this space. The action of the model consists of the
gravitational part ${\cal S}_g$ and the matter action ${\cal S}_m$
as
\begin{equation}\label{B}
{\cal S}={\cal S}_g+{\cal S}_m.\end{equation}The matter part of
the action is independent of the massive corrections to the
gravity part. Also, the gravity part can be expressed in terms of
the usual Einstein-Hilbert, with an additional correction term
coming from the massive graviton; that is \cite{Amico}
\begin{equation}\label{B1}
{\cal S}_g=\int \sqrt{-g}\left[R-\frac{m^2}{4}{\cal
U}(g,H)\right]d^4x,\end{equation}in which all of the modifications
due to the mass and also the interactions between the tensor
fields $H_{\mu\nu}$ and $g_{\mu\nu}$ are summarized in the
potential ${\cal U}(g,H)$. By using ghost-free conditions for the
theory in Ref. \cite{Arkani}, we propose the following form for
the potential term: \cite{Clau}
\begin{equation}\label{B2}
{\cal U}(g,H)=-4\left({\cal L}_2+\alpha_3 {\cal L}_3+\alpha_4
{\cal L}_4\right),\end{equation}where

\begin{eqnarray}\label{B3}
\left\{
\begin{array}{ll}
{\cal L}_2=\frac{1}{2}\left(<{\cal K}>^2-<{\cal K}^2>\right),\\\\
{\cal L}_3=\frac{1}{6}\left(<{\cal K}>^3-3<{\cal K}><{\cal K}^2>+2<{\cal K}^3>\right),\\\\
{\cal L}_4=\frac{1}{24}\left(<{\cal K}>^4-6<{\cal K}>^2<{\cal
K}^2>+3<{\cal K}^2>^2+8<{\cal K}><{\cal K}^3>-6<{\cal
K}^4>\right),
\end{array}
\right.\end{eqnarray}in which the tensor ${\cal K}_{\mu\nu}$ is
defined as
\begin{equation}\label{B4}
{\cal
K}^{\mu}_{\nu}(g,H)=\delta^{\mu}_{\nu}-\sqrt{\eta_{ab}\partial^{\mu}\phi^a\partial_{\nu}\phi^b},\end{equation}and
the notations $<{\cal K}>=g^{\mu\nu}{\cal K}_{\mu\nu}$, $<{\cal
K}^2>=g^{\alpha \beta}g^{\mu\nu}{\cal K}_{\alpha \mu}{\cal
K}_{\beta \nu}$,... are used for the corresponding traces. Now,
equations (\ref{B1})-(\ref{B4}) describe the gravitational part of
the action for a massive gravity theory. Since its explicit form
directly depends on the choice of scalar fields $\phi^a(x)$, it is
appropriate to concentrate on this point first. Interesting forms
for such fields should involve terms which would describe a
suitable coordinate transformation on the Minkowski space-time. In
a flat FRW background, for instance, one may select $\phi^0=f(t)$
and $\phi^i=x^i$, as is used in \cite{Amico}. Here, for the open
FRW metric (\ref{A}), we use the following ansatz proposed in
\cite{Emir}
\begin{equation}\label{B5}
\phi^0=f(t)\sqrt{1+|K|x_ix^i},\hspace{0.5cm}\phi^i=\sqrt{|K|}f(t)x^i.\end{equation}Upon
substitution of these scalar fields and also the definition of the
Ricci scalar into the relations (\ref{B1})-(\ref{B4}), we are led
to a point-like form for the gravitational Lagrangian in the
minisuperspace $\{N,a,f\}$ as
\begin{equation}\label{C}
{\cal L}_g=-\frac{3 a \dot{a}^2}{N}-3|K|Na+m^2\left(L_2+\alpha_3
L_3+\alpha_4 L_4\right),\end{equation}where
\begin{eqnarray}\label{D}
\left\{
\begin{array}{ll}
L_2=3a\left(a-\sqrt{|K|}f\right)\left(2Na-a \dot{f}-N\sqrt{|K|}f\right),\\\\
L_3=\left(a-\sqrt{|K|}f\right)^2\left(4Na-3a\dot{f}-N\sqrt{|K|}f\right),\\\\
L_4=\left(a-\sqrt{|K|}f\right)^3\left(N-\dot{f}\right),
\end{array}
\right.
\end{eqnarray}in which an overdot represents differentiation with respect to the time parameter $t$. It is seen that
this Lagrangian does not involve $\dot{N}$, which means that the
momentum conjugate to this variable vanishes. In the usual
canonical formalism of general relativity, we know this issue as
the primary constraint in the sense that the variable $N$ is not a
dynamical variable but a Lagrange multiplier in the Hamiltonian
formalism. On the other hand, Lagrangian (\ref{C}) seems to show
an additional constraint related to the St\"{u}ckelberg scalars
whose dynamics are encoded in the function $f(t)$. We see that in
spite of the common Lagrangians in which the first derivative of
the configuration variables are of second order, $\dot{f}$ appears
linearly in the Lagrangian (\ref{C}). Therefore, by computing the
momentum conjugate to $f$; that is, $P_f=\frac{\partial {\cal
L}_g}{\partial \dot{f}}$, we obtain
\begin{equation}\label{D1}
P_f=-m^2(a-\sqrt{|K|}f)\left[3a^2+3\alpha_3
a(a-\sqrt{|K|}f)+\alpha_4
(a-\sqrt{|K|}f)^2\right].\end{equation}Now, it is clear that this
relation is not invertible to obtain $\dot{f}(f,P_f)$. In such a
case, the Lagrangian is said to be singular and the relations like
(\ref{D1}), which hinder the inversion, are known as primary
constraints. One may use the method of Lagrange multipliers to
analyze the dynamics of the system by adding to the Lagrangian all
of the primary constraints multiplied by arbitrary functions of
time. However, to deal with our constrained system, we act
differently and proceed as follows. We vary the Lagrangian
(\ref{C}) with respect to $f$ to obtain
\begin{equation}\label{D2}
\left(\dot{a}-\sqrt{|K|}N\right)\left[|K|(\alpha_3+\alpha_4)f^2(t)-2\sqrt{|K|}(1+2\alpha_3+\alpha_4)a(t)f(t)+(3+3\alpha_3+\alpha_4)a^2(t)\right]=0.\end{equation}
The solution $\dot{a}=\sqrt{|K|}N$ of this equation is nothing but
what we obtain from the variation of the usual Einstein-Hilbert
Lagrangian with respect to $N$. Since its counterpart in massive
gravity is
\begin{equation}\label{D3}
\dot{a}=\frac{\sqrt{3|K|(\alpha_3+\alpha_4)^2+m^2a^2(t)\left[2(1+\alpha_3+\alpha_3^2-\alpha_4)^{3/2}-(1+\alpha_3)(2+
\alpha_3+2\alpha_3^2-3\alpha_4)\right]}}{\sqrt{3}(\alpha_3+\alpha_4)}N,\end{equation}we
cannot accept the relation $\dot{a}=\sqrt{|K|}N$ as a physical
solution. Therefore, the constraint corresponding to the dynamic
of $f(t)$ shows itself in the equation
\begin{equation}\label{D4}
\left[|K|(\alpha_3+\alpha_4)f^2(t)-2\sqrt{|K|}(1+2\alpha_3+\alpha_4)a(t)f(t)+(3+3\alpha_3+\alpha_4)a^2(t)\right]=0,\end{equation}where
using the same notation as in \cite{Emir}, its solutions can be
written as
\begin{equation}\label{E}
f(t)=\frac{X_{\pm}}{\sqrt{|K|}}a(t)\Rightarrow
\dot{f}=\frac{X_{\pm}}{\sqrt{|K|}}\dot{a},\hspace{0.5cm}X_{\pm}\equiv
\frac{1+2\alpha_3+\alpha_4 \pm
\sqrt{1+\alpha_3+\alpha_3^2-\alpha_4}}{\alpha_3+\alpha_4}.\end{equation}As
is argued in \cite{Emir}, in the limit where $\alpha_3$ and
$\alpha_4$ are of the order of a small quantity $\epsilon$, the
expression of $X_{+}$ goes to infinity while $X_{-}\rightarrow
3/2$. Because of this limiting behavior, we use the subscript $-$
in the following for numerical values of constants with subscript
$\pm$. Now we may insert the constraints (\ref{E}) into the
relations (\ref{D}) to reduce the degrees of freedom of the system
and obtain a minimal number of dynamical variables. If we do so,
we obtain
\begin{eqnarray}\label{F}
\left\{
\begin{array}{ll}
L_2=3\left(1-X_{\pm}\right)\left[\left(2-X_{\pm}\right)N-\frac{X_{\pm}}{\sqrt{|K|}}\dot{a}\right]a^3,\\\\
L_3=\left(1-X_{\pm}\right)^2\left[\left(4-X_{\pm}\right)N-3\frac{X_{\pm}}{\sqrt{|K|}}\dot{a}\right]a^3,\\\\
L_4=\left(1-X_{\pm}\right)^3\left[N-\frac{X_{\pm}}{\sqrt{|K|}}\dot{a}\right]a^3,
\end{array}
\right.
\end{eqnarray}in terms of which the Lagrangian (\ref{C}) takes its
reduced form with only one physical degree of freedom $a$. The
momentum conjugate to $a$ is
\begin{equation}\label{F}
P_a=\frac{\partial {\cal L}_g}{\partial
\dot{a}}=-\frac{6a\dot{a}}{N}+m^2\left(\frac{\partial
L_2}{\partial \dot{a}}+\alpha_3 \frac{\partial L_3}{\partial
\dot{a}}+\alpha_4 \frac{\partial L_4}{\partial
\dot{a}}\right).\end{equation}Noting that
\begin{equation}\label{G}
\frac{\partial L_2}{\partial
\dot{a}}=3\frac{X_{\pm}}{\sqrt{|K|}}(X_{\pm}-1)a^3,\hspace{0.5cm}\frac{\partial
L_3}{\partial
\dot{a}}=3\frac{X_{\pm}}{\sqrt{|K|}}(X_{\pm}-1)^2a^3,\hspace{0.5cm}\frac{\partial
L_4}{\partial
\dot{a}}=\frac{X_{\pm}}{\sqrt{|K|}}(X_{\pm}-1)^3a^3,\end{equation}
one gets
\begin{equation}\label{H}
P_a=-6\frac{a\dot{a}}{N}-\frac{C_{\pm}m^2}{\sqrt{|K|}}a^3,\hspace{0.5cm}C_{\pm}\equiv
X_{\pm}(1-X_{\pm})\left[3+3\alpha_3(1-X_{\pm})+\alpha_4
(1-X_{\pm})^2\right].\end{equation} Now, the Hamiltonian of the
model can be obtained from its standard definition
$H=\dot{a}P_a-{\cal L}$, with result
\begin{equation}\label{I}
H_g=N{\cal
H}_g=N\left[-\frac{1}{12a}\left(P_a+\frac{C_{\pm}m^2}{\sqrt{|K|}}a^3\right)^2+3|K|a+c_{\pm}m^2a^3\right],\end{equation}
in which we have defined
\begin{equation}\label{J}
c_{\pm}=\left(X_{\pm}-1\right)\left[3(2-X_{\pm})+\alpha_3(1-X_{\pm})(4-X_{\pm})+\alpha_4(1-X_{\pm})^2\right].\end{equation}We
see that the lapse function enters in the Hamiltonian as a
Lagrange multiplier as expected. Thus, when we vary the
Hamiltonian with respect to $N$, we get ${\cal H}_g=0$, which is
called the Hamiltonian constraint. On a classical level this
constraint is equivalent to the Friedmann equation, wherein our
problem at hand can be easily checked by comparing it with the
equation of motion (4.5) in \cite{Emir}. On a quantum level, on
the other hand, the operator version of this constraint
annihilates the wave function of the corresponding universe,
leading to the so-called Wheeler-DeWitt equation.

Now, let us deal with the matter field with which the action of
the model is augmented. As we have mentioned, the matter part of
the action is independent of modifications due to the mass terms.
Therefore, the matter may come into play in a common way and the
total Hamiltonian can be made by adding the matter Hamiltonian to
the gravitational part of (\ref{I}). To do this, we consider a
perfect fluid whose pressure $p$ is linked to its energy density
$\rho$ by the equation of state
\begin{equation}\label{K}
p=\omega \rho,\end{equation}where $-1\leq \omega \leq 1$ is the
equation of the stated parameter. According to Schutz's
representation for the perfect fluid \cite{Schutz}, its
Hamiltonian can be viewed as (see \cite{Vak} for details)
\begin{equation}\label{L}
H_m=N\frac{P_T}{a^{3\omega}},\end{equation}where $T$ is a
dynamical variable related to the thermodynamical parameters of
the perfect fluid and $P_T$ is its conjugate momentum. Finally, we
are in a position in which can write the total Hamiltonian
$H=H_g+H_m$ as
\begin{equation}\label{M}
H=N{\cal
H}=N\left[-\frac{1}{12a}\left(P_a+\frac{C_{\pm}m^2}{\sqrt{|K|}}a^3\right)^2+3|K|a+c_{\pm}m^2a^3+\frac{P_T}{a^{3\omega}}\right].\end{equation}
The setup for constructing the phase space and writing the
Lagrangian and Hamiltonian of the model is now complete. In the
following section, we shall deal with classical and quantum
cosmologies which can be extracted from a theory with the
previously mentioned Hamiltonian.
\section{Cosmological dynamics: classical point of view}
The classical dynamics are governed by the Hamiltonian equations.
To achieve this purpose, we divide this section into two parts. We
first consider the case in which the matter is absent, i.e., the
vacuum, and then include the matter.
\subsection{The vacuum classical cosmology}
\begin{figure}
\begin{tabular}{c}\epsfig{figure=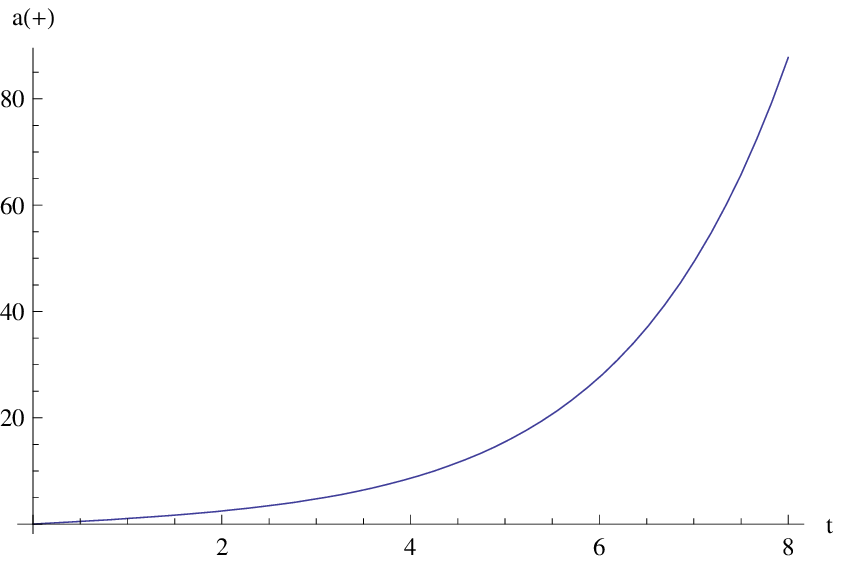,width=6cm}
\hspace{1.5cm} \epsfig{figure=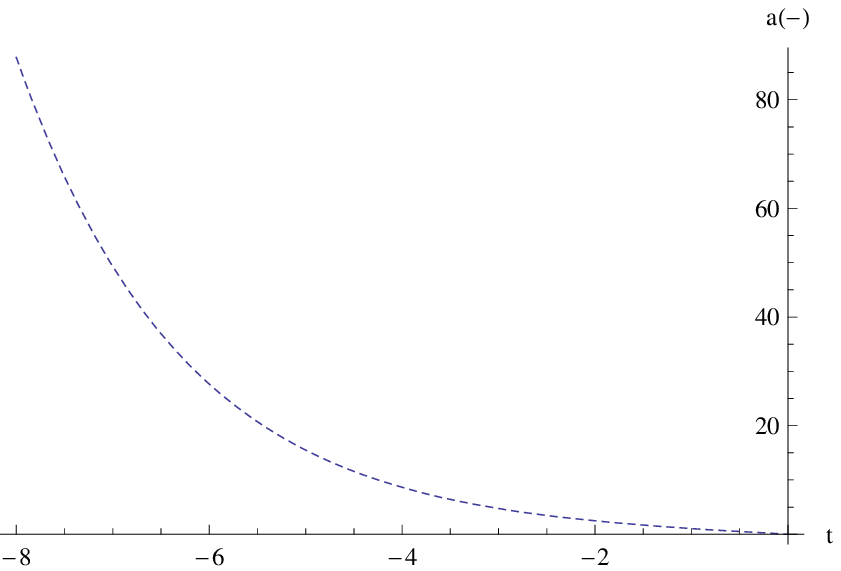,width=6cm}
\end{tabular}
\caption{\footnotesize  The figures show the evolutionary behavior
of the universes based on (\ref{P}). We have used the numerical
values $\Lambda=1$ and $\tau_{*}=0$.}\label{fig1}
\end{figure}

In this case, we can construct the equations of motion by the
Hamiltonian equations with use of the Hamiltonian (\ref{I}).
Equivalently, one may directly write the Friedmann equation from
the Hamiltonian constraint $H=0$ which, as we mentioned
previously, reflects the fact that the corresponding gravitational
theory is a parameterized theory in the sense that its action is
invariant under time reparameterization. Noting from (\ref{H})
that
\begin{equation}\label{N}
\dot{a}=-\frac{N}{6a}\left(P_a+\frac{C_{\pm}m^2}{\sqrt{|K|}}a^3\right),\end{equation}equation
(\ref{I}) gives
\begin{equation}\label{O}
3a\dot{a}^2-3|K|a=c_{\pm}m^2a^3,\end{equation}in which we have
chosen the gauge $N=1$, so that the time parameter $t$ becomes the
cosmic time $\tau$. As is indicated in \cite{Emir}, this equation
looks like the Friedmann equation for the open FRW universe with
an effective cosmological constant $\Lambda_{\pm}=c_{\pm}m^2$ and
admits the following solutions
\begin{equation}\label{P}
a_{\pm}(\tau)=\sqrt{\frac{3}{\Lambda}}\sinh \left(\pm
\sqrt{\frac{\Lambda}{3}}(\tau-\tau_{*})\right),\end{equation}
where $\tau_{*}$ is an integration constant and we have taken
$\Lambda=c_{-}m^2$. For a positive $\tau_{*}$, the condition
$a(\tau)\geq 0$ implies that the expressions of $a_{+}(\tau)$ and
$a_{-}(\tau)$ are valid for $\tau\geq \tau_{*}$ and $\tau\leq
-\tau_{*}$ respectively, such that $a_{\pm}(\tau_{*})=0$. It is
seen that the evolution of the corresponding universe with the
scale factor $a_{+}(\tau)$ begins with a big-bang-like singularity
at $\tau=\tau_{*}$ and then follows an exponential law expansion
at late time of cosmic evolution in which the mass term shows
itself as a cosmological constant. For a universe with the scale
factor $a_{-}(\tau)$, on the other hand, the behavior is opposite.
The universe decreases its size from large values of scale factor
at $\tau=-\infty$ and ends its evolution at $\tau=-\tau_{*}$ with
a zero size. In figure \ref{fig1} we have plotted theses scale
factors for typical values of the parameters. As this figure
shows, although the behavior of $a_{+}(\tau)$ ($a_{-}(\tau)$) is
like a de Sitter ($a(\tau)\sim e^{\sqrt{\Lambda/3}\tau}$) universe
at $\tau\rightarrow \infty$ ($\tau\rightarrow -\infty$), in spite
of the de Sitter, it begins (ends) its evolution with a
singularity. In summary, what we have shown previously is that in
the framework of an open FRW background geometry, the vacuum
solutions of the massive theory are equivalent to the solutions of
the usual GR with a cosmological constant. Accordingly, the
zero-size singularity of both theories has the same nature. In
this sense we would like to emphasize that the metric (\ref{A})
with the scale factor (\ref{P}) is indeed a section of the de
Sitter hyperboloid
\begin{equation}\label{P1}
-T^2+X^2+Y^2+Z^2+W^2=1,\end{equation} embedded in a
$5$-dimensional Minkowski space
\begin{equation}\label{P2}
ds^2=-dT^2+dX^2+dY^2+dZ^2+dW^2.\end{equation}To see this, one may
parameterize the hyperboloid in terms of the spherical coordinates
$(r,\theta, \phi)$ as \cite{Her}

\begin{eqnarray}\label{P3}
\left\{
\begin{array}{ll}
T=\sqrt{1+r^2}\sinh \tau,\\\\
X=\cosh \tau,\\\\Y=r\sinh \tau \cos \phi \cos \theta,\\\\Z=r\sinh
\tau \cos \phi \sin \theta,\\\\W=r\sinh \tau \sin \phi,
\end{array}
\right.
\end{eqnarray} which, upon substitution into the metric (\ref{P2}),
yields the open FRW metric with the scale factor $a(\tau)=\sinh
\tau$. This means that the point $a=0$ can be viewed as a
coordinate singularity. However, we have to note that in the
presence of any kind of matter field the point $a(\tau_{*})=0$
represents a true singularity. Thus, our following analysis to
quantize the model is based on the minisuperspace coordinate
system in terms of which the dynamical representation of the
metric, i.e. (\ref{A}), is written.

In the next section, we shall see how the previous picture may be
modified when one takes into account quantum mechanical
considerations.

\subsection{Perfect fluid classical cosmology}
Now, we assume that a perfect fluid in its Schutz's representation
is coupled with gravity. In this case the Hamiltonian (\ref{M})
describes the dynamics of the system. The equations of motion for
$T$ and $P_T$ read as
\begin{equation}\label{Q}
\dot{T}=\{T,H\}=\frac{N}{a^{3\omega}},\hspace{0.5cm}\dot{P_{T}}=\{P_T,H\}=0.\end{equation}A
glance at the above equations shows that with choosing the gauge
$N=a^{3\omega}$, we shall have
\begin{equation}\label{R}
N=a^{3\omega}\Rightarrow T=t,\end{equation}which means that
variable $T$ may play the role of time in the model. Therefore,
the Friedmann equation $H=0$ can be written in the gauge
$N=a^{3\omega}$ as follows
\begin{equation}\label{S}
3\dot{a}^2=3|K|a^{6\omega}+\Lambda
a^{6\omega+2}+P_0a^{3\omega-1},\end{equation}where we take
$P_T=P_0=\mbox{const.}$ from the second equation of (\ref{Q}).
Since it is not possible to find the analytical solutions of the
above differential equation for any arbitrary $\omega$, we present
its solutions only in some special cases.

$\bullet$ $\omega=-\frac{1}{3}$: cosmic string. In this case we
obtain
\begin{equation}\label{T}
a(t)=\left[\frac{\Lambda}{3}(t-t_0)^2-\frac{P_0+3|K|}{\Lambda}\right]^{1/2},\end{equation}where
$t_0$ is an integration constant. We see that the evolution of the
universe based on (\ref{T}) has big-bang-like singularities at
$t=t_0\pm t_{*}$ where $t_{*}=\frac{\sqrt{3(P_0+3|K|)}}{\Lambda}$.
Indeed, the condition $a^2(t)\geq 0$ separates two sets of
solutions $a^{(I)}(t)$ and $a^{(II)}(t)$, each of which is valid
for $t\leq t_0-t_{*}$ and $t\geq t_0+t_{*}$, respectively. For the
former, we have a contracting universe which decreases its size
according to a power law relation and ends its evolution in a
singularity at $t=t_0-t_{*}$, while for the latter, the evolution
of the universe begins with a big-bang singularity at
$t=t_0+t_{*}$ and then follows the power law expansion at late
time of cosmic evolution.

One may translate these results in terms of the cosmic time
$\tau$. Using its relationship with the time parameter $t$ in this
case, that is, $d\tau=a^{-1}(t)dt$, we are led to
\begin{eqnarray}\label{T1}
a(\tau)=\left\{
\begin{array}{ll}
a^{(I)}(\tau)=\frac{1}{\sqrt{12\Lambda}}\left[e^{-\sqrt{\frac{\Lambda}{3}}(\tau-\tau_0)}-3(P_0+3|K|)e^{\sqrt{\frac{\Lambda}{3}}(\tau-\tau_0)}\right]
,\hspace{.5cm}\tau \leq \tau_0-\tau_{*},\\\\\\
a^{(II)}(\tau)=\frac{1}{\sqrt{12\Lambda}}\left[e^{\sqrt{\frac{\Lambda}{3}}(\tau-\tau_0)}-3(P_0+3|K|)e^{-\sqrt{\frac{\Lambda}{3}}(\tau-\tau_0)}\right],\hspace{.5cm}
\tau \geq \tau_0+\tau_{*},\\
\end{array}
\right.
\end{eqnarray}where
$\tau_{*}=\frac{1}{2}\sqrt{\frac{3}{\Lambda}}\ln 3(P_0+3|K|)$.
Again, it is seen that there is a classically forbidden region
$\tau_0-\tau_{*}<\tau<\tau_0+\tau_{*}$, for which we have no valid
classical solutions. For $\tau \leq \tau_0-\tau_{*}$, the universe
has a exponential decreasing behavior which ends its evolution in
a singular point with zero size at $\tau=\tau_0-\tau_{*}$, while
in the region $\tau \geq \tau_0+\tau_{*}$ it begins with the
big-bang singularity at $\tau=\tau_0+\tau_{*}$ and then grows
exponentially forever.

$\bullet$ $\omega=-1$: cosmological constant.  Performing the
integration, we get the following implicit relation between $t$
and $a(t)$:
\begin{equation}\label{U}
\frac{1}{\sqrt{3}(P_0+\Lambda)^2}\left[-6|K|+(P_0+\Lambda)a^2\right]\sqrt{3|K|+(P_0+\Lambda)a^2}=t-t_0.\end{equation}In
terms of the cosmic time $\tau$, it is easy to see that this
solution returns to (\ref{P}), in which the cosmological term is
replaced by $\Lambda \rightarrow \Lambda+\mbox{cons}.$ This is
expected because the solutions (\ref{P}) were equivalent to an
open FRW universe with a cosmological constant. Therefore, adding
a new cosmological term (a perfect fluid with $\omega=-1$) only
makes a shift in the corresponding cosmological constant.

\section{Cosmological dynamics: quantum point of view}
In this section we look for the quantization of the model
presented above via the method of canonical quantization. As is
well known, this procedure is based on the Wheeler-DeWitt equation
$\hat{{\cal H}}\Psi=0$, where $\hat{{\cal H}}$ is the operator
version of the Hamiltonian constraint and $\Psi$ is the wave
function of the universe, a function of the $3$-geometries and the
matter fields. As in the case of the classical cosmology, we
consider the matter of free and perfect fluid quantum cosmology
separately.

Before going to the subject, a remark is in order related to the
Hamiltonians (\ref{I}) and (\ref{M}). The term in the round
bracket in these Hamiltonians is like the Hamiltonian of a charged
particle moving in an electromagnetic field. From this analogy,
one may define the transformation
\begin{equation}\label{U1}
P_a\rightarrow
\Pi_a=P_a+\frac{C_{\pm}m^2}{\sqrt{|K|}}a^3,\hspace{0.5cm}a\rightarrow
a,\end{equation}to simplify the form of the classical Hamiltonian.
It is clear that this is a canonical transformation both
classically and quantum mechanically \cite{And}. Since going back
from a new set of variables to the old ones in a classical
canonical transformation can be made without any ambiguity,
applying this transformation may not be important for the
classical dynamics presented in the previous section. In the
context of quantum mechanics, on the other hand, the subject is of
little difference. The transition to the quantum version of the
theory is achieved by promoting observables to operators which are
not necessarily commuting. Thus, by replacing the canonical
variables $(a,P_a)$ by their operator counterparts
$(\hat{a},\hat{P_a}=-id/da)$, we obtain the quantum Hamiltonian
\begin{equation}\label{U2}
\hat{{\cal
H}}=-\frac{1}{12}\hat{a}^{-1}\hat{\Pi_a}^2+...=-\frac{1}{12}\hat{a}^{-1}\left(\hat{P_a}+\frac{C_{\pm}m^2}{\sqrt{|K|}}\hat{a}^3\right)^2+...,\end{equation}where
$...$ denotes the terms out of the round bracket in expressions
(\ref{I}) or (\ref{M}). When calculating the square, it should be
noted that the operators $\hat{a}$ and $\hat{P_a}$ do not commute.
Although the order of these operators does not matter in the
classical analysis, quantum mechanically this issue is quite
crucial. Indeed, this is the operator ordering problem and,
unfortunately, there is no well defined principle which specifies
the order of operators in the passage from classical to quantum
theory. There are, however, some simple rules which one uses
conventionally. If, for instance, we order the products of
$\hat{a}$ and $\hat{P_a}$ in $\hat{\Pi_a}^2$ such that the
momentum stands to the right of the scale factor, we obtain
\begin{equation}\label{U3}
\hat{\Pi_a}^2\rightarrow
\hat{P_a}^2+\frac{C_{\pm}^2m^4}{|K|}a^6+2\frac{C_{\pm}m^2}{\sqrt{|K|}}\hat{a}^3\hat{P_a}-3i\frac{C_{\pm}m^2}{\sqrt{|K|}}\hat{a}^2,\end{equation}in
which we have used the commutation relation
$[\hat{a},\hat{P_a}]=i$. With this expression at hand, there is
still another factor ordering ambiguity in the terms
$\hat{a}^{-1}\hat{P_a}^2$ and $\hat{a}^2\hat{P_a}$ to construct
the quantum Hamiltonian (\ref{U2}). As Hawking and Page have shown
\cite{Haw}, the choice of different factor ordering will not
affect semiclassical calculations in quantum cosmology, so for
convenience one usually chooses a special place for it in the
special models. However, in general, the behavior of the wave
function depends on the chosen factor ordering \cite{Ste}. In what
follows, as one usually does in the minisuperspace approximation
to the cosmological models, we work in the framework of a special
factor ordering in which, in addition to the expression (\ref{U3})
for $\hat{\Pi_a}^2$, we also use the orderings
$\hat{a}^{-1}\hat{P_a}^2=\hat{P_a}\hat{a}^{-1}\hat{P_a}$ and
$\hat{a}^2\hat{P_a}=\hat{a}\hat{P_a}\hat{a}$ to make the
Hamiltonian hermitian\footnote{With the canonical transformation
(\ref{U1}) at hand, one may uses the transformed Hamiltonian
${\cal H}=-\frac{1}{12a}\Pi_a^2+...,$ to quantize the system,
where again ... denotes the terms out of the round bracket in
expressions (\ref{I}) or (\ref{M}). Using this Hamiltonian in the
hermitian form $a^{-1}\Pi_a^2=\Pi_a a^{-1}\Pi_a$ and also
representing $\Pi_a$ by $-i\partial_a$, this is equivalent to our
above treatment in which the last term in (\ref{U3}) is absent.
Therefore, one may have some doubts on the validity of the main
following results due to the effects of the chosen factor
ordering. To overcome this problem, we have made some calculations
based on the above mentioned transformed Hamiltonian and have
verified that the general patterns of the resulting wave functions
follow the behavior shown in following sections.}.
\subsection{The  vacuum quantum cosmology}
\begin{figure}
\begin{tabular}{c}\epsfig{figure=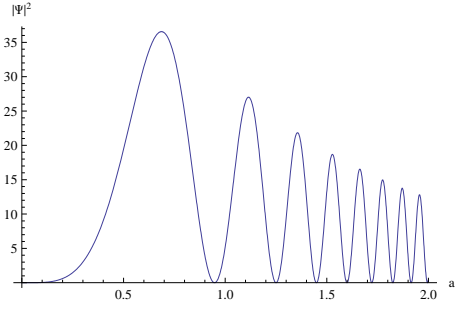,width=6cm}
\end{tabular}
\caption{\footnotesize  The square of the wave function for the
quantum universe. We take the numerical value
$\Lambda=1.5$.}\label{fig2}
\end{figure}

In this case, with the help of the Hamiltonian (\ref{I}) and use
of the abovementioned choice of ordering, the Wheeler-DeWitt
equation reads
\begin{equation}\label{V}
\left\{\frac{d^2}{da^2}+\left(-a^{-1}+2i\frac{C_{\pm}}{c_{\pm}}\Lambda
a^3\right)\frac{d}{da}+
\left[\left(36+2i\frac{C_{\pm}}{c_{\pm}}\Lambda
\right)a^2+12\Lambda
a^4-\frac{C_{\pm}^2}{c_{\pm}^2}\Lambda^2a^6\right]
\right\}\Psi(a)=0.\end{equation}This equation does not seem to
have analytical solutions. However, we can get some properties of
its solutions in special regions where there is interest in
classical and quantum regimes. First of all, let us rewrite this
equation in the form
\begin{equation}\label{X}
\left\{\frac{d^2}{da^2}-\left(a^{-1}+6i\Lambda
a^3\right)\frac{d}{da}+
\left[\left(36-6i\Lambda\right)a^2+12\Lambda a^4-9\Lambda^2
a^6\right] \right\}\Psi(a)=0,\end{equation}in which we have used
the numerical values $C_{-}=-9/4$ and $c_{-}=3/4$ \cite{Emir}. For
large values of $a$, the solution to this equation can easily be
obtained in the Wentzel-Kramers-Brillouin (WKB) (semiclassical)
approximation. In this regime we can neglect the term $a^{-1}$ in
equation (\ref{X}). Then, substituting
$\Psi(a)=\Omega(a)e^{iS(a)}$ in this equation leads to the
modified Hamilton-Jacobi equation
\begin{equation}\label{Y}
-\left(\frac{dS}{da}\right)^2+6\Lambda a^3
\frac{dS}{da}+\left(36a^2+12\Lambda a^4-9\Lambda^2
a^6\right)+{\cal Q}=0,\end{equation}in which the quantum potential
is defined as ${\cal Q}=\frac{1}{\Omega}\frac{d^2\Omega}{da^2}$.
It is well-known that the quantum effects are important for small
values of the scale factor and in the limit of the large scale
factor can be neglected. Therefore, in the semiclassical
approximation region we can omit the ${\cal Q}$ term in (\ref{Y})
and obtain
\begin{equation}\label{Z}
\frac{dS}{da}=3\Lambda a^3\pm a \sqrt{36+12\Lambda
a^2}.\end{equation} In the WKB method, the correlation between
classical and quantum solutions is given by the relation
$P_a=\frac{\partial S}{\partial a}$. Thus, using the definition of
$P_a$ in (\ref{H}), the equation for the classical trajectories
becomes
\begin{equation}\label{AB}
\dot{a}=\pm \sqrt{1+\frac{\Lambda}{3}a^2},\end{equation}from which
one finds
\begin{equation}\label{AC}
a(t)=\sqrt{\frac{3}{\Lambda}}\sinh \left(\pm
\sqrt{\frac{\Lambda}{3}}(t-\delta)\right),\end{equation}which
shows that the late time behavior of the classical cosmology
(\ref{P}) is exactly recovered. The meaning of this result is that
for large values of the scale factor, the effective action
corresponding to the expanding and contracting universes is very
large and the universe can be described classically. On the other
hand, for small values of the scale factor we cannot neglect the
quantum effects, and the classical description breaks down. Since
the WKB approximation is no longer valid in this regime, one
should go beyond the semiclassical approximation. In the quantum
regime, if we neglect the term $\Lambda^2 a^6$ in (\ref{X}), the
two linearly independent solutions to this equation can be
expressed in terms of the Hermite $H_{\nu}(x)$ and hypergeometric
$F_{1\hspace{-.5cm}1}\hspace{.4cm}(a,b;z)$ functions, leading to
the following general solution:
\begin{equation}\label{AD}
\Psi(a)=e^{-ia^2}\left[c_1H_{-\frac{1}{2}-\frac{8}{3\Lambda}i}\left(\frac{(1+i)(2+3\Lambda
a^2)}{2\sqrt{3\Lambda}}\right)+c_2 \,\,\,
F_{1\hspace{-.5cm}1}\hspace{.4cm}\left(\frac{1}{4}+\frac{4}{3\Lambda}i,\frac{1}{2};\frac{i(2+3\Lambda
a^2)^2}{6\Lambda}\right)\right].\end{equation}At this step we take
a quick glance at the question of the boundary conditions on the
solutions to the Wheeler-DeWitt equation. Note that the
minisuperspace of the above model has only one degree of freedom
denoted by the scale factor $a$ in the range $0<a<\infty$.
According to \cite{Vil}, its nonsingular boundary is the line
$a=0$, while at the singular boundary this variable is infinite.
Since the minisuperspace variable is restricted to the
abovementioned domain, the minisuperspace quantization deals only
with wave functions defined on this region. Therefore, to
construct the quantum version of the model, one should take into
account this issue. This is because in such cases, one usually has
to impose boundary conditions on the allowed wave functions;
otherwise the relevant operators, especially the Hamiltonian, will
not be self-adjoint. The condition for the Hamiltonian operator
$\hat{{\cal H}}$ associated with the classical Hamiltonian
function (\ref{I}) and (\ref{M}) to be self-adjoint is
$(\psi_1,\hat{{\cal H}}\psi_2)=(\hat{{\cal H}}\psi_1,\psi_2)$ or
\begin{eqnarray}\label{AD2}
\int_0^\infty \psi_1^*(a)\hat{{\cal H}}\psi_2(a)da= \int_0^\infty
\psi_2(a)\hat{{\cal H}}\psi_1^*(a)da.\end{eqnarray} Following the
calculations in \cite{Lem} and dealing only with square integrable
wave functions, this condition yields a vanishing wave function at
the nonsingular boundary of the minisuperspace. Hence, we impose
the boundary condition on the solutions (\ref{AD}) such that at
the nonsingular boundary (at $a=0$), the wave function vanishes.
This makes the Hamiltonian hermitian and self-adjoint and can
avoid the singularities of the classical theory, i.e. there is
zero probability for observing a singularity corresponding to
$a=0$.\footnote{Such a boundary condition is also suggested by
DeWitt in the form $\Psi[{\cal G}^{(3)}]=0$ \cite{Dew}, where
${\cal G}^{(3)}$ denotes all three-geometries which may play the
roll of barriers, for instance singular three-geometries. As is
argued in \cite{Dew}, with this boundary condition some kinds of
classical singularities can be removed and a unique solution to
the Wheeler-DeWitt equation may be obtained. Although in the
presence of more fundamental proposals of the boundary condition
in quantum cosmology (for example, Vilenkin's tunneling or
Hawking's no boundary proposals), it is not clear that the above
mentioned boundary condition is true, there are some evidences in
quantum gravity models in which suitable wave packets obey such
kind of boundary condition, see \cite{Kief}.} Therefore, we
require
\begin{equation}\label{AD1}
\Psi(a=0)=0\Rightarrow
\frac{c_2}{c_1}=-\frac{H_{-\frac{1}{2}-\frac{8}{3\Lambda}i}\left(\frac{1+i}{\sqrt{3\Lambda}}\right)}{F_{1\hspace{-.5cm}1}\hspace{.4cm}
\left(\frac{1}{4}+\frac{4i}{3\Lambda},\frac{1}{2};\frac{2i}{3\Lambda}\right)}.\end{equation}Note
that equation (\ref{X}) is a Schr\"{o}dinger-like equation for a
fictitious particle with zero energy moving in the field of the
superpotential with the real part $U(a)=-(36a^2+12\Lambda a^4)$.
Usually, in the presence of such a potential the minisuperspace
can be divided into two regions, $U>0$ and $U<0$, which could be
termed the classically forbidden and classically allowed regions,
respectively. In the classically forbidden region the behavior of
the wave function is exponential, while in the classically allowed
region the wave function behaves oscillatorily. In the quantum
tunneling approach \cite{Vil}, the wave function is so constructed
as to create a universe emerging from {\it nothing} by a tunneling
procedure through a potential barrier in the sense of usual
quantum mechanics. Now, in our model, the superpotential is always
negative, which means that there is no possibility of tunneling
anymore, since a zero energy system is always above the
superpotential. In such a case, tunneling is no longer required as
classical evolution is possible. As a consequence the wave
function always exhibits oscillatory behavior. In figure
\ref{fig2}, we have plotted the square of the wave functions for
typical values of the parameters. It is seen from this figure that
the wave function has a well-defined behavior near $a=0$ and
describes a universe emerging out of nothing without any
tunneling. (See \cite{Coul}, in which such wave functions also
appeared in the case study of the probability of quantum creation
of compact, flat, and open de Sitter universes.) On the other
hand, the emergence of several peaks in the wave function may be
interpreted as a representation of different quantum states that
may communicate with each other through tunneling. This means that
there are different possible universes (states) from which the
present universe could have evolved and tunneled in the past, from
one universe (state) to another.
\subsection{Perfect fluid quantum cosmology}
\begin{figure}
\begin{tabular}{c}\epsfig{figure=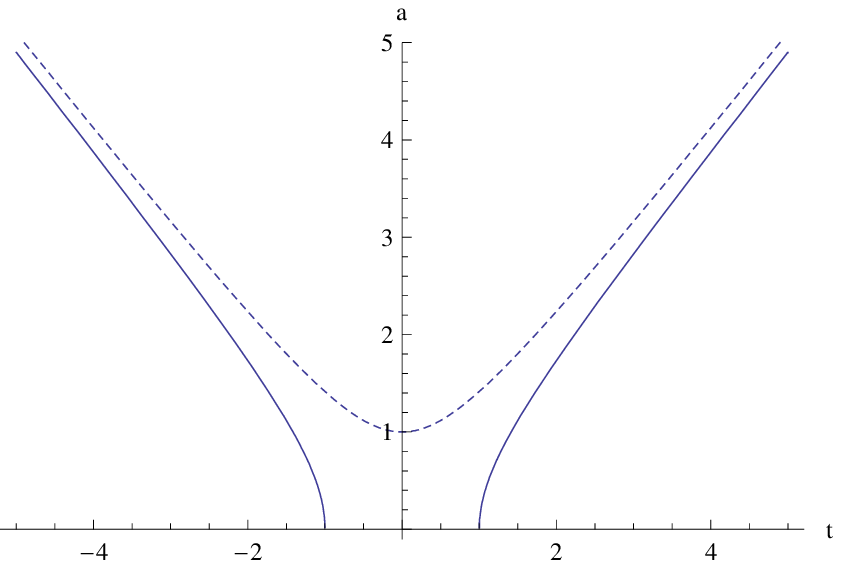,width=6cm}
\hspace{1.5cm} \epsfig{figure=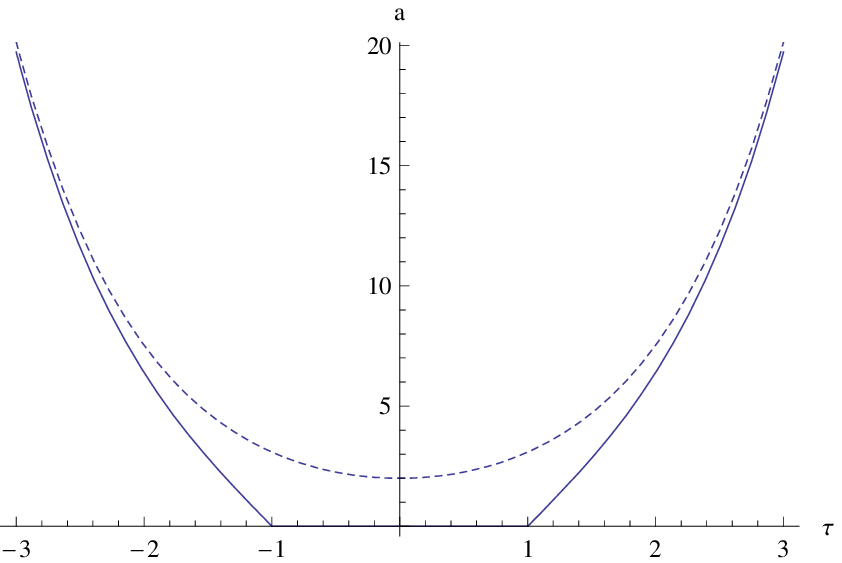,width=6cm}
\end{tabular}
\caption{\footnotesize  Left: The figure shows qualitative
behavior of the classical scale factor (\ref{T}) (solid lines: the
left branch for $a^{(I)}(t)$ and the right branch for
$a^{(II)}(t)$) and the expectation value of the scale factor
(\ref{AS}) (dashed line). Right: The same figure in terms of
cosmic time. The left and right branches of the solid lines
represent $a^{(I)}(\tau)$ and $a^{(II)}(\tau)$, respectively, in
(\ref{T1}) while the dashed line represents the expectation value
(\ref{AT}).}\label{fig3}
\end{figure}

In this case, the Wheeler-DeWitt equation can be constructed by
means of the Hamiltonian (\ref{M}). With the same approximations
as we used in the previous subsection, we obtain
\begin{equation}\label{AE}
\left\{\frac{\partial^2}{\partial a^2}-\left(a^{-1}+6i\Lambda
a^3\right)\frac{\partial}{\partial a}+
\left[\left(36-6i\Lambda\right)a^2+12\Lambda
a^4\right]-ia^{1-3\omega}\frac{\partial}{\partial T}
\right\}\Psi(a,T)=0.\end{equation}We separate the variables in
this equation as
\begin{equation}\label{AF}
\Psi(a,T)=e^{iET}\psi(a),\end{equation}leading to
\begin{equation}\label{AG}
\left\{\frac{d^2}{d a^2}-\left(a^{-1}+6i\Lambda
a^3\right)\frac{d}{d a}+
\left[\left(36-6i\Lambda\right)a^2+12\Lambda
a^4+Ea^{1-3\omega}\right]\right\}\psi(a)=0.\end{equation}The
solutions of the above differential equation may be written in the
form \begin{equation}\label{AH}
\psi_E(a)=e^{-ia^2}\left[c_1H_{-\frac{1}{2}-\frac{32+E}{12\Lambda}i}\left(\frac{(1+i)(2+3\Lambda
a^2)}{2\sqrt{3\Lambda}}\right)+c_2 \,\,\,
F_{1\hspace{-.5cm}1}\hspace{.4cm}\left(\frac{1}{4}+\frac{32+E}{24\Lambda}i,\frac{1}{2};\frac{i(2+3\Lambda
a^2)^2}{6\Lambda}\right)\right],\end{equation}for $\omega=-1/3$
and
\begin{eqnarray}\label{AI}
\psi_E(a)=e^{-i(1+\frac{E}{12
\Lambda})a^2}\left[c_1H_{-\frac{1}{2}-\frac{1152\Lambda^2-24E\Lambda-E^2}{432\Lambda^3}i}\left(\frac{(1+i)\left[E+6\Lambda(2+3\Lambda
a^2)\right]}{12\Lambda\sqrt{3\Lambda}}\right)+\nonumber \right.\\
\left.
c_2\,\,\,F_{1\hspace{-.5cm}1}\hspace{.4cm}\left(\frac{1}{4}+\frac{1152\Lambda^2-24E\Lambda-E^2}{864\Lambda^3}i,\frac{1}{2};\frac{i\left[E+6\Lambda(2+3\Lambda
a^2)\right]^2}{216\Lambda^3}\right)\right],\end{eqnarray}for
$\omega=-1$. Now the eigenfunctions of the Wheeler-DeWitt equation
can be written as
\begin{equation}\label{AJ}
\Psi_E(a,T)=e^{iET}\psi_E(a).\end{equation}We may now write the
general solution to the Wheeler-DeWitt equation as a superposition
of its eigenfunctions; that is,
\begin{equation}\label{AK}
\Psi(a,T)=\int_0^\infty A(E)\Psi_E(a,T)dE,\end{equation}where
$A(E)$ is a suitable weight function to construct the wave
packets. The above relations seem to be too complicated to extract
an analytical expression for the wave function. Therefore, in the
following (for the case $\omega=-1/3$), we present an approximate
analytic method which is valid for very small values of scale
factor, i.e., in the range that we expect the quantum effects to
be important. In this regime if we keep only the $a^{-1}$ and
$a^2$ terms in the second and third terms of (\ref{AG}), the
solutions to this equation can be viewed as a superposition of the
functions $\sin\left(\frac{\sqrt{36+E-6i\Lambda}}{2}a^2\right)$
and $\cos\left(\frac{\sqrt{36+E-6i\Lambda}}{2}a^2\right)$. If we
impose the boundary condition $\psi(a=0)=0$ on these solutions, we
are led to the following eigenfunctions:
\begin{equation}\label{AL}
\Psi_E(a,T)=e^{iET}\sin\left(\frac{\sqrt{36+E-6i\Lambda}}{2}a^2\right).\end{equation}
Now, by using the equality
\begin{equation}\label{AM}
\int_0^\infty e^{-\gamma x}\sin \sqrt{mx}dx=\frac{\sqrt{\pi
m}}{2\gamma^{3/2}}e^{-(m/4\gamma)},\end{equation}we can evaluate
the integral over $E$ in (\ref{AK}), and the simple analytical
expression for this integral is found if we choose the function
$A(E)$ to be a quasi-Gaussian weight factor $A({\cal
E})=e^{-\gamma {\cal E}}$ ($\gamma$ is an arbitrary positive
constant and ${\cal E}=36+E-6i\Lambda$), which results in
\begin{equation}\label{AN}
\Psi(a,T)=e^{-6(\Lambda+6i)T}\int_0^\infty e^{-\gamma {\cal
E}}e^{i{\cal E}T}\sin\left(\frac{\sqrt{{\cal
E}}}{2}a^2\right)d{\cal E}.\end{equation}Using the relation
(\ref{AM}) yields the following expression for the wave function
\begin{equation}\label{AO}
\Psi(a,T)={\cal
N}e^{-6(\Lambda+6i)T}\frac{a^2}{(\gamma-iT)^{3/2}}\exp
\left(-\frac{a^2}{8(\gamma-iT)}\right),\end{equation}where ${\cal
N}$ is a numerical factor. Now, having this expression for the
wave function of the universe, we are going to obtain the
predictions for the behavior of the dynamical variables in the
corresponding cosmological model. To do this, one may calculate
the time dependence of the expectation value of a dynamical
variable $q$ as
\begin{equation}\label{AP}
<q>(T)=\frac{<\Psi|q|\Psi>}{<\Psi|\Psi>}.\end{equation}Following
this approach, we may write the expectation value for the scale
factor as
\begin{equation}\label{AR}
<a>(T)=\frac{\int_0^\infty
\Psi^{*}(a,T)a\Psi(a,T)da}{\int_0^\infty
\Psi^{*}(a,T)\Psi(a,T)da},\end{equation}which yields
\begin{equation}\label{AS}
<a>(T)=\sqrt{\frac{\Lambda}{3}}\left(\gamma^2+T^2\right)^{1/2}.\end{equation}This
relation may be interpreted as the quantum counterpart of the
classical solutions (\ref{T}). However, in spite of the classical
solutions, for the wave function (\ref{AO}), the expectation value
(\ref{AS}) of $a$ never vanishes, showing that these states are
nonsingular. Indeed, in (\ref{AS}) $T$ varies from $-\infty$ to
$+\infty$, and any $T_0$ is just a specific moment without any
particular physical meaning like big-bang singularity. The above
result may be written in terms of the cosmic time $\tau$. By the
definition $d\tau=a^{-1}(T)dT$, we obtain the quantum version of
the relations (\ref{T1}) as
\begin{equation}\label{AT}
<a>(\tau)=\frac{1}{2}\left(
e^{\sqrt{\frac{\Lambda}{3}}\tau}+\gamma^2
e^{-\sqrt{\frac{\Lambda}{3}}\tau}\right).\end{equation} In figure
\ref{fig3}, we have plotted the classical scale factors (\ref{T})
and (\ref{T1}) and their quantum counterparts (\ref{AS}) and
(\ref{AT}). As is clear from this figure, for a perfect fluid with
$\omega=-1/3$, the corresponding classical cosmology admits two
separate solutions which are disconnected from each other by a
classically forbidden region. One of these solutions represents a
contracting universe ending in a singularity while another
describes an expanding universe which begins its evolution with a
big-bang singularity. On the other hand, the evolution of the
scale factor based on the quantum-mechanical considerations shows
a bouncing behavior in which the universe bounces from a
contraction epoch to a reexpansion era. Indeed, the classically
forbidden region is where the quantum bounce has occurred. We see
that in the late time of cosmic evolution in which the quantum
effects are negligible, these two behaviors coincide with each
other. This means that the quantum structure which we have
constructed has a good correlation with its classical counterpart.

\section{Bohmian trajectories}
\begin{figure}
\begin{tabular}{c}\epsfig{figure=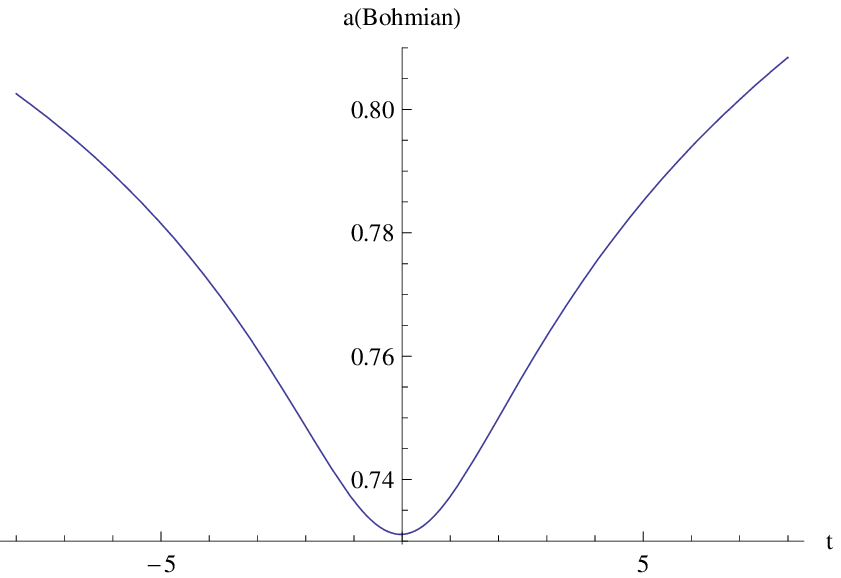,width=6cm}
\hspace{1.5cm} \epsfig{figure=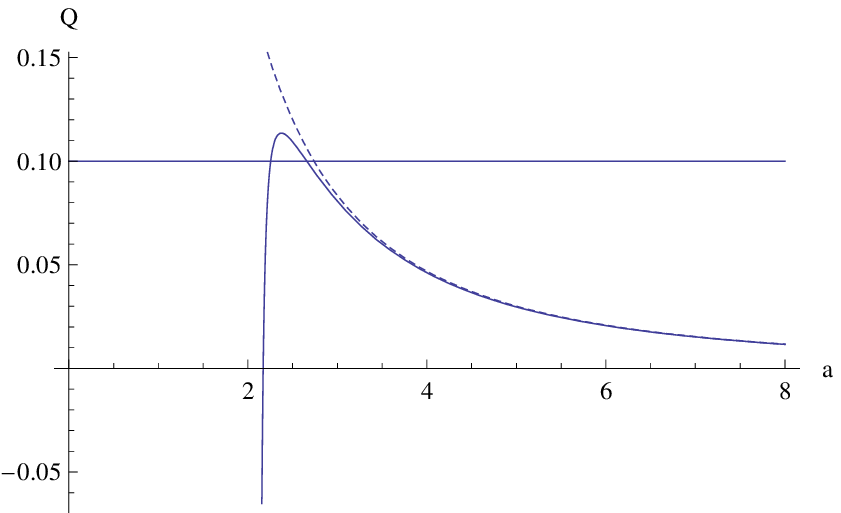,width=6cm}
\end{tabular}
\caption{\footnotesize  Left: The Bohmian trajectory of the scale
factor. Right: The horizontal line represents a typical energy
level. The solid curve is the quantum potential for $\Lambda \neq
0$, while the dashed curve denotes the quantum potential when
$\Lambda=0$.}\label{fig4}
\end{figure}

In the previous sections, we saw how the classical singular
behavior of the universe was replaced with a bouncing one in a
quantum picture. Now, a natural question may arise: Why will the
bounce occur? Clearly, it is due to the quantum mechanical effects
which show themselves when the size of the universe tends to very
small values. However, we would like to know whether the massive
correction to the underlying gravity theory has any contribution
to this phenomenon. To deal with this question, let us return to
the wave function (\ref{AO}) and write it in the polar form
$\Psi(a,T)=\Omega(a,T)e^{iS(a,T)}$, where $\Omega(a,T)$ and
$S(a,T)$ are real functions, which simple algebra gives as
\begin{equation}\label{AU}
\Omega(a,T)=e^{-6\Lambda
T}\frac{a^2}{(\gamma^2+T^2)^{3/4}}\exp\left[-\frac{\gamma
a^2}{8(\gamma^2+T^2)}\right],\end{equation}
\begin{equation}\label{AV}
S(a,T)=-36T+\frac{3}{2}\arctan
\frac{T}{\gamma}-\frac{Ta^2}{8(\gamma^2+T^2)}.\end{equation}According
to the Bohm-de Broglie interpretation of quantum mechanics
\cite{Bohm} and also its usage in quantum cosmology \cite{Pin},
upon using this form of the wave function in the corresponding
wave equation, we arrive at the modified Hamilton-Jacobi equation
as
\begin{equation}\label{AX}
{\cal H}\left(q_i,P_i=\frac{\partial S}{\partial q_i}\right)+{\cal
Q}=0,\end{equation}where $P_i$ are the momentum conjugate to the
dynamical variables $q_i$ and ${\cal Q}$ is the quantum potential.
With beginning of the wave equation (\ref{AE}), for which we have
used the same approximations as in the previous section, the above
mentioned procedure gives the quantum potential as
\begin{equation}\label{AY}
{\cal Q}=\frac{1}{\Omega}\frac{\partial^2 \Omega}{\partial
a^2}-\frac{1}{a\Omega}\frac{\partial \Omega}{\partial
a}.\end{equation}On the other hand, the Bohmian equations of
motion can be obtained by $P_a=\frac{\partial S}{\partial a}$,
where by means of the relation (\ref{H}) reads
\begin{equation}\label{AZ}
-6a\dot{a}+3\Lambda
a^2=-\frac{T}{4(\gamma^2+T^2)}.\end{equation}The solution to this
equation denotes the Bohmian representation of the scale factor;
that is
\begin{equation}\label{BA}
a(t)=\sqrt{ce^{\Lambda T}+\frac{1}{24}e^{\Lambda T-i\gamma
\Lambda}\left[e^{2i\gamma \Lambda}\mbox{ Ei}(1;-\Lambda T-i\gamma
\Lambda)+\mbox{Ei}(1;-\Lambda T+i\gamma
\Lambda)\right]},\end{equation}where $c$ is an integration
constant and $\mbox{Ei}(b;z)$ is the exponential integral function
defined by
\begin{equation}\label{BC}
\mbox{Ei}(b;z)=\int_1^\infty e^{-kz}k^{-b}dk.\end{equation}The
bouncing behavior of the scale factor is again its main property
near the classical singularities as we have shown in figure
\ref{fig4}. To achieve an expression for the quantum potential in
terms of the scale factor, we note that all of our above
calculations are in the vicinity of $T\sim 0$, where the scale
factor is small. In this regime, a numerical analysis shows that
the Bohmian scale factor (\ref{BA}) behaves as $a(T)\sim
(\gamma^2+T^2)^{1/2}$, in agreement with the expectation value
(\ref{AS}). Thus, substituting in (\ref{AU}), we get the quantum
potential from (\ref{AY}) as
\begin{equation}\label{BD}
{\cal Q}(a)=\frac{3}{4}\left[\gamma^2 \left(\frac{1}{a^4 -
\gamma^2 a^2}+\frac{8\Lambda}{(a^2 - \gamma^2)^{3/2}}\right)+
\frac{-1+48 \Lambda ^2 a^2}{a^2-\gamma^2}\right].\end{equation}

In figure \ref{fig4} we also have plotted the qualitative
behavior of the quantum potential versus the scale factor. As this
figure shows, this potential goes to zero for the large values of
the scale factor. This behavior is expected, since in this regime
the quantum effects can be neglected and the universe evolves
classically. On the other hand, for the small values of the scale
factor the potential takes a large magnitude and the quantum
mechanical considerations come into the scenario. This is where
the quantum potential can produce a huge repulsive force which may
be interpreted as being responsible of the avoidance of
singularity. In figure \ref{fig4} the horizontal line represents a
constant energy level which in intersecting with the potential
curves gives the turning points at which the bounce will occur.
The solid curve in this figure is plotted in the case of $\Lambda
\neq 0$; i.e., for the massive theory, while the dashed curve is
for $\Lambda=0$; i.e., for when the massive corrections are
absent. It is seen that, although the mass term $\Lambda$ is not
the only reason for the bouncing behavior in the vicinity of the
classical singularity, it may shift the bouncing point into the
smaller values of the scale factor. This means that if we consider
the bouncing point as the minimum size of the universe (which is
suggested by quantum cosmology), then the massive version of the
underlying gravity theory predicts a smaller value for this
minimal size in comparison with the usual Einstein-Hilbert model.
These facts and also other considerable possibilities such as
quantum tunneling between different classically allowed regimes
(as can be seen from figure \ref{fig4}) through the potential
barrier support the idea that the massive corrections to the
classical cosmology are some signals from quantum gravity.

\section{Conclusions}
In this paper we have applied the recently proposed nonlinear
massive theory of gravity to an open FRW cosmological setting.
Although the absence of homogeneous and isotropic solutions is one
of the main challenges related to this kind of gravitational
theory, we moved along the lines of \cite{Emir, Emir1}, in which
the existence of open FRW cosmologies is investigated. By using
the constraint corresponding to the St\"{u}ckelberg scalars, we
reduced the number of degrees of freedom, according to which the
total Hamiltonian of the model is deduced. We then presented in
detail, the classical cosmological solutions either for the empty
universe or in the case where the universe is filled by a perfect
fluid (in its Schutz representation) with the equation of state
parameter $\omega=-1/3,-1$. We saw that in both of these cases,
the solutions consist of a contraction universe which finalizes
its evolution in a singular point and an expanding universe which
begins its dynamic with a big-bang singularity. These two branches
of solutions are disconnected from each other by a classically
forbidden region. Also, the common feature of the vacuum and
matter classical solutions is that the mass term plays a role
which resembles the role of cosmological constant in the usual de
Sitter universe. In this sense we may relate the massive
corrections of GR to the problem of dark energy.

In another part of the paper, we dealt with the quantization of
the model described above via the method of canonical
quantization. For an empty universe, we have shown that by
applying the WKB approximation on the Wheeler-DeWitt equation, one
can recover the late time behavior of the classical solutions. For
the early universe, we obtained oscillatory quantum states free of
classical singularities by which two branches of classical
solutions may communicate with each other. In the presence of
matter, we focused our attention on the approximate analytical
solutions to the Wheeler-DeWitt equation in the domain of small
scale factor, i.e. in the region which the quantum cosmology is
expected to be dominant. Using Schutz's representation for the
perfect fluid, under a particular gauge choice, we led to the
identification of a time parameter which allowed us to study the
time evolution of the resulting wave function. Investigation of
the expectation value of the scale factor shows a bouncing
behavior near the classical singularity. In addition to
singularity avoidance, the appearance of bounce in the quantum
model is also interesting in its nature due to prediction of a
minimal size for the corresponding universe. We know the idea of
existence of a minimal length in nature is supported by almost all
candidates of quantum gravity. Finally, we repeated the quantum
calculations by means of the Bohmian approach to quantum
mechanics. The analysis of the quantum potential shows the
importance of the mass term in the action of the model. Indeed, we
have shown that in the presence of the massive graviton, the
quantum potential changes its behavior from an infinite barrier to
a finite one, and hence the minimal size of the universe, from
which the bounce occurs, will be shifted to the smaller values.
Also, the massive theory of quantum cosmology exhibits some other
possibilities; for example, tunneling between different
classically allowed regions, for cosmic evolution in the early
universe epoch.


\begin{thebibliography}{99}
\bibitem{pauli} M. Fierz and W. Pauli, {\it Proc. R. Soc.} A {\bf 173} (1939) 211

\bibitem{deser} D.G. Boulware and S. Deser, {\it Phys. Rev.} D {\bf 6} (1972) 3368

\bibitem{derahm} C. de Rham and G. Gabadadze, {\it Phys. Rev.} D {\bf 82} (2010) 4 (arXiv: 1007.0443
[hep-th])

\bibitem{works}K. Koyama, G. Niz and G. Tasinato, {\it Phys. Rev. Lett.} {\bf 107} (2011) 131101 (arXiv: 1103.4708
[hep-th])\\K. Koyama, G. Niz and G. Tasinato, {\it Phys. Rev.} D
{\bf 84} (2011) 064033 (arXiv: 1104.2143 [hep-th])\\T.M.
Nieuwenhuizen, {\it Phys. Rev.} D {\bf 84} (2011) 024038 (arXiv:
1103.5912 [gr-qc])\\S.F. Hassan and R.A. Rosen, {\it Phys. Rev.
Lett.} {\bf 108} (2012) 041101 (arXiv: 1106.3344 [hep-th])\\C. de
Rham, G. Gabadadze and A.J. Tolley, {\it Phys. Lett.} B {\bf 711}
(2012) 190 (arXiv: 1107.3820 [hep-th])\\C. de Rham, G. Gabadadze
and A.J. Tolley, {\it Helicity Decomposition of Ghost-free Massive
Gravity}, (arXiv: 1108.4521 [hep-th])\\L. Berezhiani, G.
Chkareuli, C. de Rham, G. Gabadadze, A.J. Tolley, {\it On Black
Holes in Massive Gravity} (arXiv: 1111.3613 [hep-th])

\bibitem{Amico}G. D'Amico, C. de Rham, S. Dubovsky, G. Gabadadze,
D. Pirtskhalava and A.J. Tolley, {\it Phys. Rev.} D {\bf 84}
(2011) 124046 (2011) (arXiv: 1108.5231 [hep-th])

\bibitem{Emir}A.E. G\"{u}mr\"{u}k\c{c}\"{u}o\u{g}lu, C. Lin and S.
Mukohyama, {\it J. Cosmol. Astropart. Phys.} {\bf 11} (2011) 030
(arXiv: 1109.3845 [hep-th])

\bibitem{Emir1} A.E. G\"{u}mr\"{u}k\c{c}\"{u}o\u{g}lu, C. Lin and S.
Mukohyama, {\it J. Cosmol. Astropart. Phys.} {\bf 03} (2012) 006
(arXiv: 1111.4107 [hep-th])

\bibitem{nima}M.S. Volkov, {\it J. High Energy Phys.} {\bf 01} (2012) 035 (arXiv: 1110.6153
[hep-th]),\\ D. Comelli, M. Crisostomi, F. Nesti and L. Pilo, {\it
J. High Energy Phys.} {\bf 03} (2012) 065 (arXiv: 1111.1983
[hep-th])\\M. von Strauss, A. Schmidt-May, J. Enander, E. Mortsell
and S.F. Hassan, \textit{Cosmological Solutions in Bimetric
Gravity and Their Observational Tests}, (arXiv: 1111.1655
[gr-qc])\\N. Khosravi, N. Rahmanpour, H.R. Sepangi and S. Shahidi,
{\it Phys. Rev.} D {\bf 85} (2012) 024049 (arXiv: 1111.5346
[hep-th])\\N. Khosravi, H.R. Sepangi and S. Shahidi, {\it On
massive cosmological scalar perturbations}, (arXiv: 1202.2767
[gr-qc])

\bibitem{hassan}S.F. Hassan and R.A. Rosen, {\it J. High Energy Phys.} {\bf 01} (2012) 126 (arXiv: 1109.3515 [hep-th])\\
S.F. Hassan and R.A. Rosen, \textit{Confirmation of the Secondary
Constraint and Absence of Ghost in Massive Gravity and Bimetric
Gravity}, (arXiv: 1111.2070 [hep-th])\\S.F. Hassan and R.A. Rosen,
\textit{On Non-Linear Actions for Massive Gravity}, (arXiv:
1103.6055 [hep-th])


\bibitem{Clau}C. de Rham, G. Gabadadze and A.J. Tolley, {\it Phys. Rev. Lett.} {\bf 106} (2011) 231101 (arXiv: 1011.1232
[hep-th])

\bibitem{Arkani}N. Arkani-Hamed, H. Georgi and M.D. Schwartz, {\it
Ann. Phys.} {\bf 305} (2003) 96 (arXiv: hep-th/0210184)\\S.L.
Dubovsky, {\it J. High Energy Phys.} {\bf 10} (2004) 076 (arXiv:
hep-th/0409124)

\bibitem{Schutz}B.F. Schutz, {\it Phys. Rev.} D {\bf 2} (1970)
2762\\B.F. Schutz, {\it Phys. Rev.} D {\bf 4} (1971) 3559\\V.G.
Lapchinskii, V.A. Rubakov, {\it Theor. Math. Phys.} {\bf 33}
(1977) 1076


\bibitem{Vak}A.B. Batista, J.C. Fabris, S.V.B. Goncalves and J. Tossa, {\it Phys. Lett.} A {\bf 283} (2001) 62
(arXiv: gr-qc/0011102)\\ F.G. Alvarenga, J.C. Fabris, N.A. Lemos
and G.A. Monerat, {\it Gen. Rel. Grav.} {\bf 34} (2002) 651
(arXiv: gr-qc/0106051)\\ A.B. Batista, J.C. Fabris, S.V.B.
Goncalves and J. Tossa, {\it Phys. Rev.} D {\bf 65} (2002) 063519
(arXiv: gr-qc/0108053)\\B. Vakili, {\it Phys. Lett.} B {\bf 688}
(2010) 129 (arXiv: 1004.0306 [gr-qc])\\B. Vakili, {\it Class.
Quantum Grav.} {\bf 27} (2010) 025008 (arXiv: 0908.0998 [gr-qc])

\bibitem{Her}{\O}. Gr{\o}n and S. Hervik, {\it
Einstein's General Theory of Relativity} (Springer, New York,
2007)

\bibitem{And} A. Anderson, {\it Ann. Phys.} {\bf 232} (1994) 292 (arXiv:
hep-th/9305054)

\bibitem{Haw}S.W. Hawking and D.N. Page, {\it Nucl. Phys.} B {\bf 264} (1986) 185

\bibitem{Ste} R. Steigl and F. Hinterleitner, {\it Class. Quantum Grav.} {\bf 23} (2006) 3879

\bibitem{Vil}A. Vilenkin, {\it Phys. Rev.} D {\bf 37} (1988) 888\\ A. Vilenkin,
{\it Phys. Rev.} D {\bf 33} (1986) 3560

\bibitem{Lem}N.A. Lemos, {\it J. Math. Phys.} {\bf 37} (1996) 1449 (arXiv: gr-qc/9511082)

\bibitem{Dew}B.S. DeWitt, {\it Phys. Rev.} {\bf 160} (1967) 1113

\bibitem{Kief}C. Kiefer, {\it Quantum Gravity} (Oxford University Press,
New York, 2007).

\bibitem{Coul}D.H. Coule and J. Martin, {\it Phys. Rev.} D {\bf 61} (2000) 063501 (arXiv: gr-qc/9905056)\\A.
Linde, {\it J. Cosmol. Astropart. Phys.} {\bf 10} (2004) 004
(arXiv: hep-th/0408164)

\bibitem{Bohm}D. Bohm, {\it Phys. Rev.} {\bf 85} (1952) 166\\
D. Bohm, {\it Phys. Rev.} {\bf 85} (1952) 180\\ P.R. Holland, {\it
The Quantum Theory of Motion: An Account of the de Broglie-Bohm
Interpretation of Quantum Mechanics}, (Cambridge University Press,
Cambridge, 1993)

\bibitem{Pin}F.T. Falciano and N. Pinto-Neto, {\it Phys. Rev.} D {\bf 79} (2009) 023507 (arXiv:
0810.3542 [gr-qc])\\A. Shojai and F. Shojai, {\it Europhys. Lett.}
{\bf 71} (2005) 886 (arXiv: gr-qc/0409020)
\end{thebibliography}
\end{document}